%% file: main.tex
\documentclass[trackchanges,twocolumn]{aastex7}

\usepackage{comment}
\usepackage{multirow} 
\usepackage{amsmath}  
\usepackage{xcolor}  
\usepackage{caption}  

\begin{document}

\title{Ion-Neutral Drift Velocity as a Diagnostic of Dust Growth and Magnetic Field in Star-Forming Environments}

\author[]{Haruka Fukihara}
\affiliation{Kagoshima University}
\email[show]{k9786351@kadai.jp}  

\author[]{Yusuke Tsukamoto}
\affiliation{Kagoshima University}
\email{}

\author[]{Hiroyuki Hirashita}
\affiliation{Osaka University}
\email{}

\author[]{Doris Arzoumanian}
\affiliation{Kyusyu University}
\email{}

\author[]{Yoshiaki Misugi}
\affiliation{National Astronomical Observatory of Japan}
\affiliation{Kyusyu Sangyo University}
\email{}

\begin{abstract}
Recent observations have revealed that the ion–neutral drift velocity in star-forming molecular clouds and dense cores is on the order of $\sim100\ \rm{m\,s^{-1}}$. 
Theoretical studies have shown that, in ambipolar diffusion —the process responsible for the differential motion between ions and neutrals— the dust size distribution has a significant impact on the magnetic resistivities. 
In this study, we perform simulations to investigate how dust growth through accretion and coagulation affects the ion–neutral drift velocity in molecular clouds and cores. 
We find that, on core scales, both dust growth and a magnetic field strength of $200\ \rm{\mu G}$ are required to reproduce the observed drift velocity. 
We suggest that measurements of ion-neutral drift velocity, particularly on core scales, may serve as a new diagnostic to constrain the dust size distribution and magnetic field strength in such environments.

\end{abstract}



\section{INTRODUCTION} 
Magnetic fields in star-forming clouds are primarily probed by Zeeman splitting and by dust polarization.
Zeeman observations in $\rm H_{I}$, OH, and CN provide direct line-of-sight field strengths in diffuse and dense gas \citep{2005ApJ...624..773H, 2008ApJ...680..457T, 2008A&A...487..247F, 2012ARA&A..50...29C}.
However, detections of Zeeman splitting require bright, suitable transitions, and high signal-to-noise. 
Therefore, the measurements are inevitably sparse and subject to selection effects. 

Dust polarization observations constrain the magnetic-field orientation in the plane of the sky.
Furthermore, magnetic field strengths can be inferred using the Davis–Chandrasekhar–Fermi (DCF) method, which relates the dispersion of polarization angles to the gas velocity dispersion \citep{1953ApJ...118..113C}.
Large surveys such as the JCMT POL-2 BISTRO project now provide cloud-to-core coverage of polarization maps \citep{2019ApJ...877...88C, 2020ApJ...899...28D,  2019ApJ...880...27P, 2021ApJ...907...88P, 2021A&A...647A..78A, 2021ApJ...913...85H, 2021ApJ...918...85L, 2022ApJ...941..122C, 2023ApJ...946...62W, 2024ApJ...962..136W}

However, DCF-based estimates may carry substantial systematics (projection, beam and line-of-sight averaging, density and turbulence assumptions, order-unity calibration factors), and recent assessments highlight non-negligible biases and uncertainties \citep{2012ARA&A..50...29C, 2021A&A...647A.186S, 2022MNRAS.514.1575C}. 
Given these complementary but limited techniques: Zeeman splitting is precise but observationally restrictive, whereas DCF method is widely applicable but uncertain, there is clear motivation to develop additional, independent diagnostics of magnetic field strength in star-forming regions.

Ambipolar diffusion, the slippage of magnetic fields through partially ionized gas mediated by ion–neutral drift, has long been recognized as a key ingredient in star formation theory \citep[e.g.,][]{1956MNRAS.116..503M,1987ARA&A..25...23S}.
By considering ambipolar diffusion, the induction equation can be written as
\begin{equation}
\frac{\partial\boldsymbol{B}}{\partial t}
=\nabla\times(\boldsymbol{v}\times\boldsymbol{B})
+\nabla\times\!\left[\eta_{\rm A}\big((\nabla\times\boldsymbol{B})\times\hat{\boldsymbol{B}}\big)\times\hat{\boldsymbol{B}}\right],
\label{eq_induction_ad}
\end{equation}
where $\eta_{\rm A}$ is the ambipolar resistivity and $\hat{\boldsymbol{B}}$ the unit vector along $\boldsymbol{B}$.
The value of $\eta_{\rm A}$ is calculated from the ionization balance, which is controlled by cosmic-ray ionization, gas-phase recombination, and the adsorption of charged particles onto dust grains.

The crucial role of dust grains in regulating the ionization degree and ambipolar resistivities was clarified by early work that explicitly treated dust–controlled ionization and coupling in molecular clouds and cores \citep[e.g.,][]{1979ApJ...232..729E, 1980PASJ...32..405U, 1986MNRAS.218..663N, 1990MNRAS.243..103U, 1991ApJ...368..181N}; see also \citet{2002ApJ...573..199N}.
Subsequently, axisymmetric thin-disk/1.5D calculations incorporating dust-dependent magnetic resistivities were applied to the dynamical evolution of molecular cloud cores \citep{1993ApJ...418..774C,1994ApJ...425..142C,1995ApJ...454..194C}.
Multi-dimensional non-ideal MHD studies further linked this microscopic physics to core collapse and disk formation, exploring the impact of Ohmic dissipation, Hall effect, and ambipolar diffusion with varying the ionization chemistry \citep[e.g.,][]{2011ApJ...738..180L, 2015MNRAS.452..278T, 2015ApJ...801..117T, 2016A&A...587A..32M, 2016MNRAS.457.1037W, 2016MNRAS.460.2050Z, 2021MNRAS.508.2142X, 2022MNRAS.510.4420T, 2023MNRAS.521.2661K, 2023PASJ...75..835T, 2024PASJ...76..674T, 2025A&A...695A..18T}.
It is commonly understood that the ionization degree, and hence $\eta_{\rm A}$, is sensitive to the abundance of small grains \citep{2021MNRAS.505.5142Z,2022ApJ...934...88T}.

On the observational side, several efforts have been made to constrain ion-neutral drift directly by comparing matched-resolution ion and neutral tracers.
\citet{2018A&A...615A..58Y} found that the ion–neutral velocity difference on $\sim 10^2$–$10^3$~AU scales in B335 is small ($\lesssim$ a few $\times 10^2~\mathrm{m\,s^{-1}}$), implying strong coupling between the magnetic field and gas on these scales.
On cloud scales, \citet{2021ApJ...912....7P} reported systematic ion–neutral linewidth differences in Barnard~5 filament (N$_2$H$^+$ broader than NH$_3$), and additional detections are emerging in a prestellar core (Arzoumanian \& Spezzano et al., in prep.).
Nevertheless, interpreting ion-neutral drift measurements requires careful consideration: line-of-sight averaging, projection, tracer chemistry/optical depth, and large-scale field geometry can bias inferred drifts \citep[e.g.,][]{2009ApJ...706.1504H, 2008ApJ...677.1151L, 2025A&A...695A..18T}.

Despite the well-established importance of dust grains for ambipolar diffusion, and hence ion-neutral drift, the explicit connection between dust size evolution and ion-neutral drift on cloud-to-core scales has received comparatively little attention.
Given the strong sensitivity of $\eta_{\rm A}$ to the small-grain population and the growing observational evidence of ion-neutral drift, it is timely to reassess how the evolution of the dust size distributions affects the ion-neutral drift velocity and how ion-neutral drift measurements can constrain magnetic fields and dust properties.

In this context, from molecular clouds to dense cores, there are several observations that suggest an evolving dust size distribution. 
The “coreshine” phenomenon: mid-IR scattered light from dense cores may require micron-sized grains \citep{2010Sci...329.1622P, 2010A&A...511A...9S, 2015A&A...582A..70S}. 
Submillimeter observations also show changes of the spectral index of dust opacity in envelope scale \citep{2009ApJ...696..841K, 2015ApJ...808..102K, 2024ApJ...961...90C}.
These results suggest possible grain growth from cloud to core, implying that gas-phase chemical reactions and hence the gas coupling to the magnetic field can change across these environments.
 
In this study, we investigate how magnetic field strength and key microphysical processes,
especially dust size evolution, affect the ion–neutral drift velocity induced by ambipolar diffusion, denoted as $v_{\rm drift}$, and how its observed values can constrain these quantities.
We consider two representative environments, a molecular cloud and a dense core, and evaluate $v_{\rm drift}$ on each scale.

This paper is organized as follows. 
In Section \ref{sec:method}, we describe our cloud and core models, the dust evolution model, and the formulation used to evaluate the ambipolar resistivity and the resulting ion–neutral drift velocity $v_{\rm drift}$. 
In Section \ref{sec:results}, we present the time evolution of the dust size distribution and the resulting $v_{\rm drift}$, and we compare our results with observations. 
In Section \ref{sec:discusion}, we discuss the implied constraints on the physical parameters from the perspective of reproducing the observed drift velocity, and we summarize our main conclusions and future prospects.

\section{METHOD} 
\label{sec:method}

\subsection{Dust evolution}
\label{subsec:method_dust_evol}
In this study, we calculated the dust size evolution considering the accretion and coagulation.

The time evolution of the dust size distribution is given by the formulation of \citet{2019MNRAS.482.2555H},
\begin{equation}
\frac{\partial\rho_d(m,\ t)}{\partial t}
=\left[\frac{\partial\rho_d(m,\ t)}{\partial t}\right]_{\rm acc}+\left[\frac{\partial\rho_d(m,\ t)}{\partial t}\right]_{\rm coag},
\end{equation}
where $\rho_d(m,\ t)$ is the mass density of the dust grains which have a mass of $m$ at a time of $t$.

For accretion, we follow the formulation presented in \citep{2012MNRAS.422.1263H}. 
\begin{equation}
    \frac{\partial n_{d}(a,\ t)}{\partial t}+\frac{\partial}{\partial a}[n_{d}(a,~t)\dot{a}]=0.
\end{equation}
\begin{equation}
    \dot{a}=\xi(t) a/\tau(a)
    \label{eq_adot}
\end{equation}
\begin{equation}
    \xi(t)\equiv n_{\rm X}(t)/n_{\rm X, tot}
\end{equation}
$\tau(a)$ in Equation~(\ref{eq_adot}) is accretion time scale and is assumed as 
\begin{equation}
    \tau(a)\equiv 
    \frac{a}{\frac{n_{X, \rm tot}m_{X}S}{f_{X}\rho_{\rm mat}}\left(\frac{k_{\rm B} T_{\rm gas}}{2\pi m_{X}}\right)^{1/2}}.
\end{equation}
where $m_X$ is the mass of element $X$.
In this study, we consider the accretion of O, which subsequently becomes $\rm H_2O$ on the grain surface.
$f_X$ is the mass fraction of the element in the dust grain material, and we adopted $f_X=16/18=0.89$.
$S$ is the sticking probability and we adopted $S=0.3$.
$k_{\rm B}$ and $T_{\rm gas}$ are Boltzmann constant and gas temperature, respectively, for which we assume $T_{\rm gas}=10\ \rm K$.
$\rho_{\rm mat}$ is the material density and we consider this as a parameter (see Section~\ref{subsec:method_model_param}).
$n_{X,\rm tot}$ is the number density of element $X$ 
and is described as
\begin{equation}
    n_{X,\rm tot}=Z_{\rm m}~x(X)~n_{\rm gas},
\end{equation}
where $Z_{\rm m}$ is metallicity and we assume $Z_{\rm m}=1\ \rm{Z_{\odot}}$.
$x(X)$ is the solar abundance of X, and we assume $x(O)=4.35\times 10^{-4}$ based on the commonly adopted interstellar value of $x(O)\sim 10^{-4}$ \citep[e.g.,][]{2018MNRAS.476.1371C}. 
With these values, dust-to-gas mass ratio becomes $1.4$ after accretion. 
$n_{\rm gas}$ is the number density of gas.

$\xi(t)$ is the fraction of the element in gas phase. 
We adopted $\xi=0.9$ at initial state and its subsequent decrease is treated as depletion according to
\begin{equation}
    \frac{\rm{d} \xi}{\rm{d} t} = 
    -\frac{3f_{\rm X}\xi(t)}{m_{\rm X}n_{\rm X,tot}}
    \int_0^\infty\frac{\sigma(m,t)}{\tau(m)}dm.
\end{equation}

The time evolution of grain size distribution by coagulation between grains with the masses of $m_{1}$ and $m_2$ (with the radius of $a_1$ and $a_2$ ) is expressed as
\begin{align}
\left[\frac{\partial\rho_d(m,\ t)}{\partial\ t}\right]_{\rm coag}
&=-m\rho_d(m,\ t)\int_0^{\infty}\alpha(m_1,\ m)\rho_d(m_1,\ t)dm_1 \notag\\
&+\int_0^{\infty}\int_0^{\infty}\alpha(m_1,\ m_2)\rho_d(m_1,\ t)\rho_d(m_2,\ t)\notag\\
&\times m_1\delta(m-m_1-m_2)dm_1dm_2.
\end{align}
Here, $\alpha$ represents the collision efficiency and is given by
\begin{equation}
    \alpha(m_1,\ m_2)\equiv\frac{\sigma_{1,2}\Delta V}{m_1m_2},
\end{equation}
where $\sigma_{1,2}=\pi(a_1+a_2)^2$ and $\Delta V$ are respectively collisional cross-section and relative velocity between two grains.

\begin{figure}[ht!]
\centering
    \includegraphics[width=8cm]{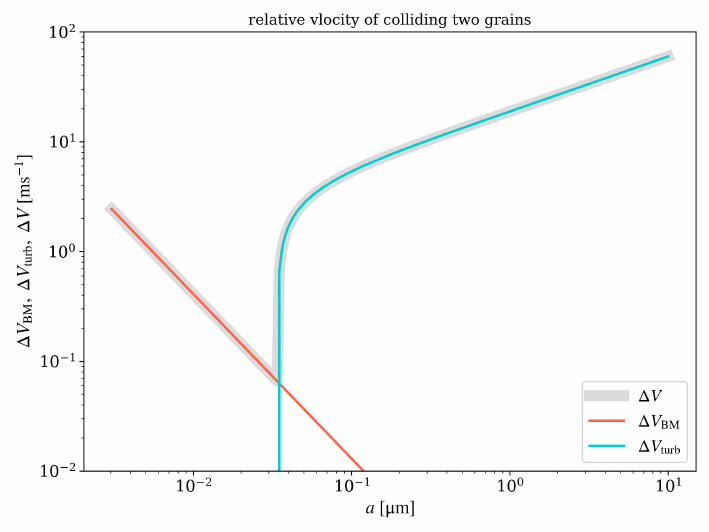}
\caption{Relative velocity model of two colliding grains introduced in this study.
The plot shows the case where the two grains have the same radius ($a_1 = a_2 = a$) and collide in a gas at $n_{\rm gas}=10^6\,\rm{cm^{-3}}$ with a Mach number of $\mathcal{M}=2$. 
$\Delta V_{\rm BM}$ represents the contribution from the Brownian motion, whereas $\Delta V_{\rm turb}$ represents the contribution from turbulent motion.
$\Delta V = \sqrt{(\Delta V_{\rm BM})^2 + (\Delta V_{\rm  turb})^2}$.}
\label{fig:rel_vel}
\end{figure}

For the relative velocity between grains, we consider contributions from Brownian motion and turbulence-induced motion, following the model of \citet{2007A&A...466..413O}.
\begin{equation}
    \Delta V = \sqrt{(\Delta V_{\rm BM})^2 + (\Delta V_{\rm turb})^2}
\end{equation}
Figure \ref{fig:rel_vel} shows the relative velocity model of two grains with the same radius in a gas of number density $n_{\rm gas}=10^6\,\rm{cm^{-3}}$ employed in this study.
We neglect the bouncing and fragmentation due to the grain-grain collision.

\subsection{Ionization chemistry and ion-neutral drift velocity calculation}
\label{subsec:method_ion_chem}

For the ionization chemistry and resistivity, we follow the analytical framework of \citet{2022ApJ...934...88T} with one modification appropriate for our cold, low-density conditions.

The ionization equilibrium equation in gas including dust grains adsorption is given as,
\begin{align}
    \zeta n_{\rm gas}-s_{\rm i}u_{\rm i}n_{\rm i}\overline{(\sigma_{d}\langle\tilde{J_{\rm i}}(I,\ Z)\rangle)}n_d-\beta n_{\rm i}n_{\rm e}=0,\\
    \zeta n_{\rm gas}-s_{\rm e}u_{\rm e}n_{\rm e}\overline{(\sigma_{d}\langle\tilde{J_{\rm e}}(I,\ Z)\rangle)}n_d-\beta n_{\rm i}n_{\rm e}=0,
\end{align}
where $n_{\rm gas}$, $n_{i}$, $n_{e}$ and $n_d$ are the number densities of the neutral gas, ion , electron and dust grains, respectively. 
$I$ and $Z$ are indices for dust size and dust charging and $\langle f \rangle$ and $\bar{f}$ denote the average over $Z$ and $I$, respectively.
The first term represents the ionization of neutral particles in the gas phase, the second term represents the adsorption of charged particles onto dust grain surfaces, and the third term represents the recombination of gas-phase ions.
$\zeta$ is the ionization rate.
$s_{i}$ and $s_{e}$ are the sticking probability of ion and electron to dust grain surface.
$u_{i}$ and $u_{e}$ are the average velocity of ion and electron.
$\tilde{J}_{i(e)}$ is the effective cross-section normalized by $\sigma_{d}(I)=\pi(a_{d}(I))^2$, where $a_{d}$ is the dust radius.
$\beta$ is the recombination rate of ions.

In the low-\(\tau\) limit (\(\tau(I)\equiv a_I k_{\rm B}T/e^2\ll1\)) appropriate for our cold, low-density conditions, the dust charge distribution concentrates to \(Z\in\{-1,0,+1\}\). Therefore, we only consider these three charge states. 
The detailed balance between adjacent dust charge states yields 
\begin{align}
\epsilon\,n_{d}(I,Z)\,\tilde{J}_i(I,Z)=n_{d}(I,Z+1)\,\tilde{J}_e(I,Z+1),
\end{align}
with \(\epsilon\equiv(n_i s_i u_i)/(n_e s_e u_e)\).

Then, charge neutrality condition is written as an equation of $\epsilon$ as,
\begin{align}
    n_i(\epsilon)-n_e(\epsilon)+n_d\overline{\langle Z\rangle}(\epsilon)=0,
\end{align}
and we solve for $\epsilon$ in this equation using the Newton-Raphson method to obtain $n_i$, $n_e$, $n_d$.

Using obtained $n_i$, $n_e$, $n_d$, the ohmic($\sigma_O$), Hall($\sigma_H$), Pedersen($\sigma_P$) conductivities are calculated as follows.
\begin{align}
    \sigma_{\rm O}=\sum_{s}{\frac{c}{B}n_{s}\beta_{s}},\\
    \sigma_{\rm H}=-\sum_{s}{\frac{c}{B}\frac{n_{s}q_{s}\beta_{s}^2}{1+\beta_{s}^2}},\\
    \sigma_{\rm P}=\sum_{s}{\frac{c}{B}\frac{n_{s}q_{s}\beta_{s}}{1+\beta_{s}^2}}.
\end{align}
The subscript $s$ denotes the particle species (ions, electrons, and dust).
$q_{s}$ is the charge of each particle species and $c$ is the speed of light.
The dimensionless Hall parameter $\beta_{s}$ is defined as the ratio of the cyclotron frequency of charged particles to their collision frequency with neutral gas particles. 
This parameter is expressed as
\begin{equation}
   \beta_s=
   \frac{q_{s} B}{m_{s}c}
   \frac{m_{s}+m_{\rm gas}}{\langle\sigma v\rangle_{s}m_{\rm gas} n_{\rm gas}},
\end{equation}
where $\langle\sigma v\rangle_{s}$ is the collisional momentum transfer rate between charged particles and neutrals and is derived from equations in \citet{2008A&A...484...17P}.
$m_{s}$ and $m_{\rm gas}$ are the mass of each particle species and the mean mass of the gas, respectively.

From the above electrical conductivities, 
the ambipolar resistivity $\eta_{\rm A}$ is calculated as,
\begin{equation}
    \eta_{\rm A} = \frac{c^2}{4\pi}\left(\frac{\sigma_{\rm P}}{\sigma_{\rm H}^2+\sigma_{\rm P}^2}-\frac{1}{\sigma_{\rm O}}\right).
\end{equation}
Then, we can estimate the ion-neutral drift velocity as,
\begin{equation}
    v_{\rm drift}=\eta_{\rm A}\frac{(\nabla\times\boldsymbol{B})\times\boldsymbol{B}}{|\boldsymbol{B}|^2}
    \sim\frac{\eta_{\rm A}}{L},
\label{eq:v_drift}
\end{equation}
where $L$ is a characteristic length scale and assumed to be the Jeans length $L=\lambda_{\rm J}$.
Note that, under the assumption that the magnetic field is frozen in the gas, it is physically natural to expect that magnetic field gradients and curvatures will arise at the Jeans length scale, which is the typical scale for the motion of self-gravitating gas.
This assumption that ion-neutral drift occurs on scales comparable to density variations (or at $\lambda_{\rm J}$ scale) has also been employed in previous studies \citep[e.g.,][]{2020A&A...641A..39S, 2024A&A...690A..23V}.

\subsection{Model and parameters}
\label{subsec:method_model_param}
In this study, we introduced a molecular cloud model and a cloud core model defined by gas density $n_{\rm gas}$ and magnetic field strength $B=|\boldsymbol{B}|$.
The molecular cloud model has $n_{\rm gas}=1.0\times10^4\ \rm{cm^{-3}}$ (corresponding mass density, free-fall time
and Jeans length is $\rho_{\rm gas}=3.91\times10^{-20}\ \rm{g\ cm}^{-3}$, $t_{\rm ff}=1.75\times10^5\ \rm yr$ and $\lambda_{\rm J}=0.27\ \rm pc$, respectively\footnote{
We calculated free-fall time as
\begin{equation}
t_{\rm ff}=\frac{1}{\sqrt{4\pi G\rho_{\rm gas}}}.
\end{equation}
Here, $\rho_{\rm gas}=\mu m_p n_{\rm gas}$ where $\mu=2.3$ is the average molecular weight for typical gas consist of ISM and $m_p=1.67\times10^{-24}\ \rm g$ is the mass of a proton.
Jeans length is given by
\begin{equation}
\lambda_{\rm J}=c_s\sqrt{\frac{\pi}{G\rho_{\rm gas}}},
\end{equation}
with $c_s=(\gamma k_{\rm B} T_{\rm gas}/m_p\mu)^{1/2}$, where $\gamma$ and $T$ is the heat capacity ratio and gas temperature, respectively. $k_{\rm B}$ is the Boltzmann constant.
}) and $B=20,\,50\ \rm{\mu G}$, while the cloud core model has $n_{\rm gas}=1.0\times10^6\ \rm{cm^{-3}}$ (corresponds to $\rho_{\rm gas}=3.91\times10^{-18}\ \rm{g\ cm}^{-3}$, $t_{\rm ff}=1.75\times10^4\ \rm yr$ and $\lambda_{\rm J}=0.03\ \rm pc$) and $B=100,  200\ \rm{\mu G}$.
We then compute dust growth and ionization chemistry in these environments (See Section \ref{subsec:method_dust_evol} and \ref{subsec:method_ion_chem}.).
The initial dust size distribution is set to the MRN distribution \citep{1977ApJ...217..425M}, with a minimum size of $a_{\rm min} =5.0\ \rm{nm}$ and a maximum size of $a_{\rm max} =250\ \rm nm$.
For the numerical implementation, the grid covers a size range from $a_{\mathrm{grid,min}} = 3\ \rm nm$ to $a_{\mathrm{grid,max}} = 30\ \rm \mu m$, discretized into 256 bins.

In the simulation, we adopted four parameters: the cosmic-ray ionization rate $\zeta_{\rm CR}$, the dominant ion species $s_{\rm ion}$, and the turbulent velocity (or Mach number $\mathcal{M}$) for the gas phase, and the material density $\rho_{\rm mat}$ for the dust phase.

The cosmic-ray ionization rate of $\zeta_{\rm CR}\sim10^{-17}\ \mathrm{s^{-1}}$ has been widely adopted in studies of star-forming regions  (so-called Spitzer value; \citet{1968ApJ...152..971S}).
Since molecular gas effectively shields cosmic rays, the ionization rate varies inversely with gas density. 
In the more diffuse Galactic ISM, values around $\sim10^{-16}\ \mathrm{s^{-1}}$ have been reported \citep{2012ApJ...745...91I}, whereas in dense molecular cloud cores, values ranging from $\sim 10^{-17}-10^{-16}\ \mathrm{s^{-1}}$ have been observed \citep{1998ApJ...499..234C}.
Therefore, in this study, we adopt $\zeta_{\rm CR} = 10^{-17},\  10^{-16}\ \mathrm{s^{-1}}$.
 
The gas velocity that determines the grain velocity is considered from subsonic to mildly supersonic ($\mathcal{M}=0.5,\ 1.0,\ 2.0$). 
Recent observations in nearby filamentary molecular clouds indicate velocity structures around the transonic regime \citep{2013A&A...553A.119A, 2011A&A...533A..34H, 2016A&A...587A..97H}.
Cloud cores that form through fragmentation of these filaments are thought to inherit subsonic velocity structures.
In fact, observations have revealed molecular cloud cores with transonic or subsonic turbulent velocities \citep{2004A&A...416..191T}. 
Theoretical studies have also shown that cloud cores can form from filaments initially possessing turbulence with $\mathcal{M}=2$ \citep{2024ApJ...963..106M}.

As the dominant gas-phase ion species, we consider $\rm HCO^+$ and $\rm H_3^+$, both of which are expected to be major ion species in molecular clouds and cloud cores.
Observations and theoretical studies (e.g., \citet{2002ApJ...565..344C, 2012ApJ...753...29T}) suggest that the $\rm{H_3^+}$ becomes the dominant ion in high-density regions ($\sim10^{6}\ \rm{cm^{-3}}$), rather than $\rm{HCO^+}$.
These ions behave differently in terms of their thermal velocity and recombination rate.
The recombination rates for $\rm HCO^+$ and $\rm H_3^+$ are taken from the UMIST database \citep{2013A&A...550A..36M}.\footnote{\url{https://umistdatabase.uk}}

For dust material, we focus on not only carbon- and silicate- based compounds but also icy grains, because the temperature of dark clouds embedding star-forming regions is sufficiently low for volatile molecules to stick to dust surface. 

The definition of models and parameters are summarized in Table \ref{tab_param}.
We take the model with $\zeta_{\rm CR}=10^{-17}$, $s_{\rm ion}=\rm{H_3^+}$, $\rho_{\rm mat}=1.0$, $v_{\rm gas}=2c_{s}(\mathcal{M}=2.0)$ as the fiducial model.

\begin{table*}[ht]
\caption{Model and parameters}
\label{tab_param}
\begin{tabular}{ccccccc}
\hline
\hline
Model & $n_{\rm gas}\ \rm{[cm^{-3}]}$ & $B\ \rm{[\mu G]}$ & $\zeta_{\rm CR}\ [\rm s^{-1}]$ & $s_{\rm ion}$ & $\rho_{\rm mat}\ \rm{[gcm^{-3}]}$ & $v_{\rm gas}$ \\
\hline
\multirow{2}{*}{Cloud}
& \multirow{2}{*}{$10^4$} & $20$
& \multirow{4}{*}{\shortstack[c]{${1.0\times10^{-17}}^*$\\$1.0\times10^{-16}$}}
& \multirow{4}{*}{\shortstack[c]{${\rm HCO^+}$\\$\rm {H_3^+}^*$}}
& \multirow{4}{*}{\shortstack[c]{$1.0^*$\\$2.0$\\$3.0$}} & \multirow{4}{*}{\shortstack[c]{$\mathcal{M}=0.5$\\$\mathcal{M}=1.0$\\$\mathcal{M}=2.0^*$}} \\
&   & $50$ &   &   &   & \\
\cline{1-3}
\multirow{2}{*}{Core}
& \multirow{2}{*}{$10^6$} & $100$ &   &   &   & \\
&   & $200$ &   &   &   & \\
\hline
\end{tabular}
\tablecomments{We assume two star-forming environment models defined by gas density and magnetic field strength, and perform one-zone calculation in each environment.
For the cloud model ($n_{\rm gas}=10^4~\rm{cm^{-3}}$), the free-fall time is $t_{\rm ff,cloud}=1.75\times10^5~\rm{yr}$ and Jeans length is $\lambda_{\rm J, cloud}=0.27~\rm{pc}$.
For the core model ($n_{\rm gas}=10^6~\rm{cm^{-3}}$), the values are $t_{\rm ff,core}=1.75\times10^4~\rm{yr}$ and $\lambda_{\rm J, core}=0.027~\rm{pc}\sim5600 ~\rm{AU}$, respectively.
We assume the temperature to be $T_{\rm gas}=10$ K.
Values marked with * are fiducial models.}
\end{table*}

\section{RESULTS} \label{sec:results}
In this section, we present the simulation results for the evolution of the dust size distribution and its impact on the ion–neutral drift velocity \(v_{\rm drift}\).

\begin{figure*}[ht!]
\includegraphics[width=0.495\linewidth]{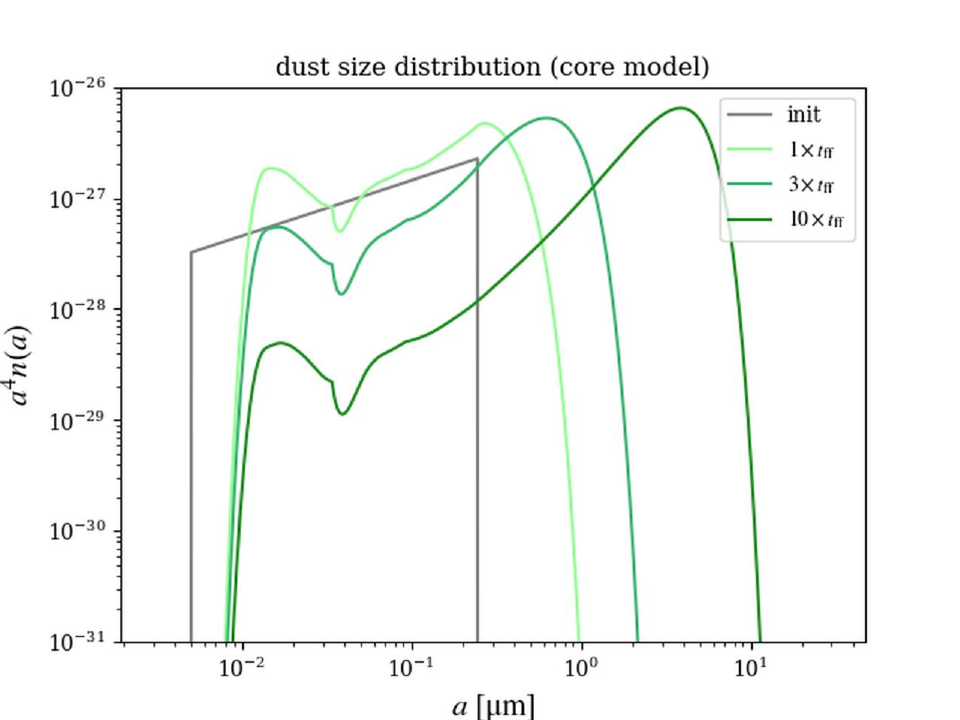}
\includegraphics[width=0.495\linewidth]{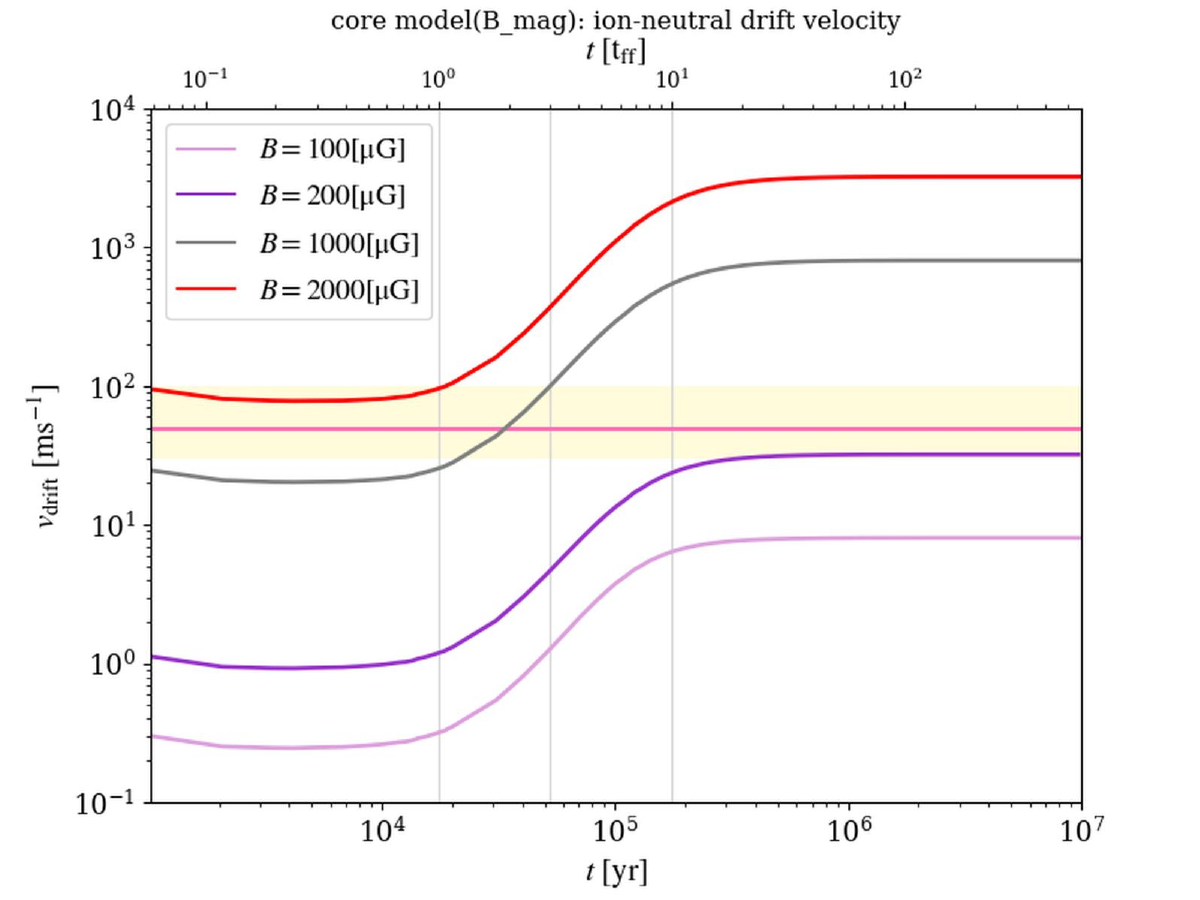}
\caption{(Left)The dust size distribution at the epoch of $1\times t_{\rm ff}$(light  green), $3\times t_{\rm ff}$(middle green), $10\times t_{\rm ff}$(deep green).
The initial size distribution is shown in gray.
(Right)Dependence of the time evolution of ion-neutral drift velocity in core model on magnetic field strength. In addition to the case of $B=100\ \rm\mu G$(light purple) and $B=200\ \rm\mu G$(purple), we plot the case of $B=1000\ \rm\mu G$(gray) and $B=2000\ \rm\mu G$(red) for reference. The lower and upper horizontal axis show the evolution time in unit of $\rm yr$ and $t_{\rm ff}$, respectively. The observed $v_{\rm drift}$ is indicated by yellow-filled region. The pink line shows $v_{\rm drift}=50\ \rm ms^{-1}$. Three thin gray vertical lines correspond to  $t=1,\ 3$ and $10\times\ t_{\rm ff}$.
\label{fig:core_vdrift_Bmag}
}
\end{figure*}

Time evolution of the dust size distribution in the fiducial core model is shown in the left panel of Figure~\ref{fig:core_vdrift_Bmag}.
Grains smaller than \(10\,\mathrm{nm}\) are efficiently depleted by accretion by \(t \simeq 1\,t_{\rm ff}\).
Subsequently, coagulation accelerates grain growth and the maximum grain size reaches \(\sim 10\,\mu\mathrm{m}\) by later times ($t\simeq 10\,t_{\rm ff}$).
A shallow dip appears around \(a \sim 30\,\mathrm{nm}\) which is attributable to the adopted relative-velocity model for colliding grains (see Figure~\ref{fig:rel_vel}).
This radius approximately marks the transition from the viscous (Brownian motion-dominated) to the inertial (turbulence–dominated) regime (see also \citet{2022MNRAS.515.2072K}).
As grains enter the inertial regime their relative velocities increase, which accelerates coagulation.

The right panel of Figure~\ref{fig:core_vdrift_Bmag} shows the time evolution of \(v_{\rm drift}\), estimated from our \(\eta_{\rm A}\) calculation via Equation~(\ref{eq:v_drift}).
The results show that the depletion of nano-sized very small grains contributing to electrical conductivity leads to an increase in the ambipolar resistivity, and consequently, a higher drift velocity.
Simultaneously, coagulation reduces the total grain surface area, weakening the adsorption of charged particles onto grains.
As a result, the gas-phase ionization degree increases and is solely determined by gas-phase recombination. Thus, both the magnetic resistivity and \(v_{\rm drift}\) increase as dust grains grow and reach a plateau at later times (for this model, \(t \gtrsim 10\,t_{\rm ff}\), i.e., \(\sim 10^6\) yr).

In this core model, the observed level of $v_{\rm drift}$ ($\sim 100 ~{\rm m s^{-1}}$) is reproduced with the magnetic  field of \(B \sim 200\,\mu\mathrm{G}\) once dust grains grow sufficiently.
Several hundred micro-Gauss is a reasonable value for the magnetic field in the cores \citep{2012ARA&A..50...29C}.

On the other hand, if grains have not grown (i.e., at the initial time of the evolution), a strong magnetic field of $2000\ \rm \mu G$ is required to explain the value of the observed drift velocity.
However, such a field strength —of several thousand micro-Gauss— is unacceptably large for the magnetic field in the cloud cores.

This result strongly suggests that dust growth is essential to produce the observed \(v_{\rm drift}\sim 100\,\mathrm{m\,s^{-1}}\) in dense cores unless extraordinarily strong magnetic fields are required.

\begin{figure*}[ht!]
\centering

\includegraphics[width=0.495\linewidth]{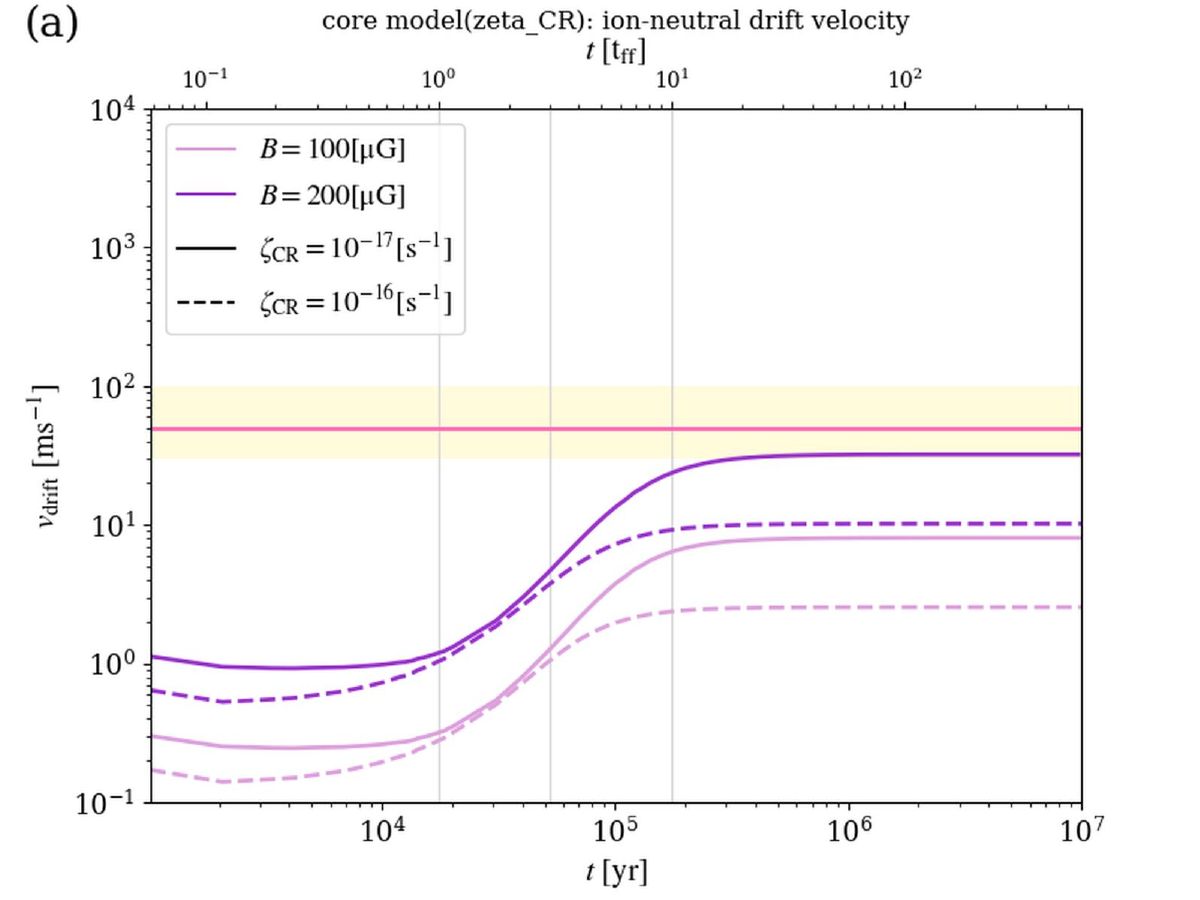}
\includegraphics[width=0.495\linewidth]{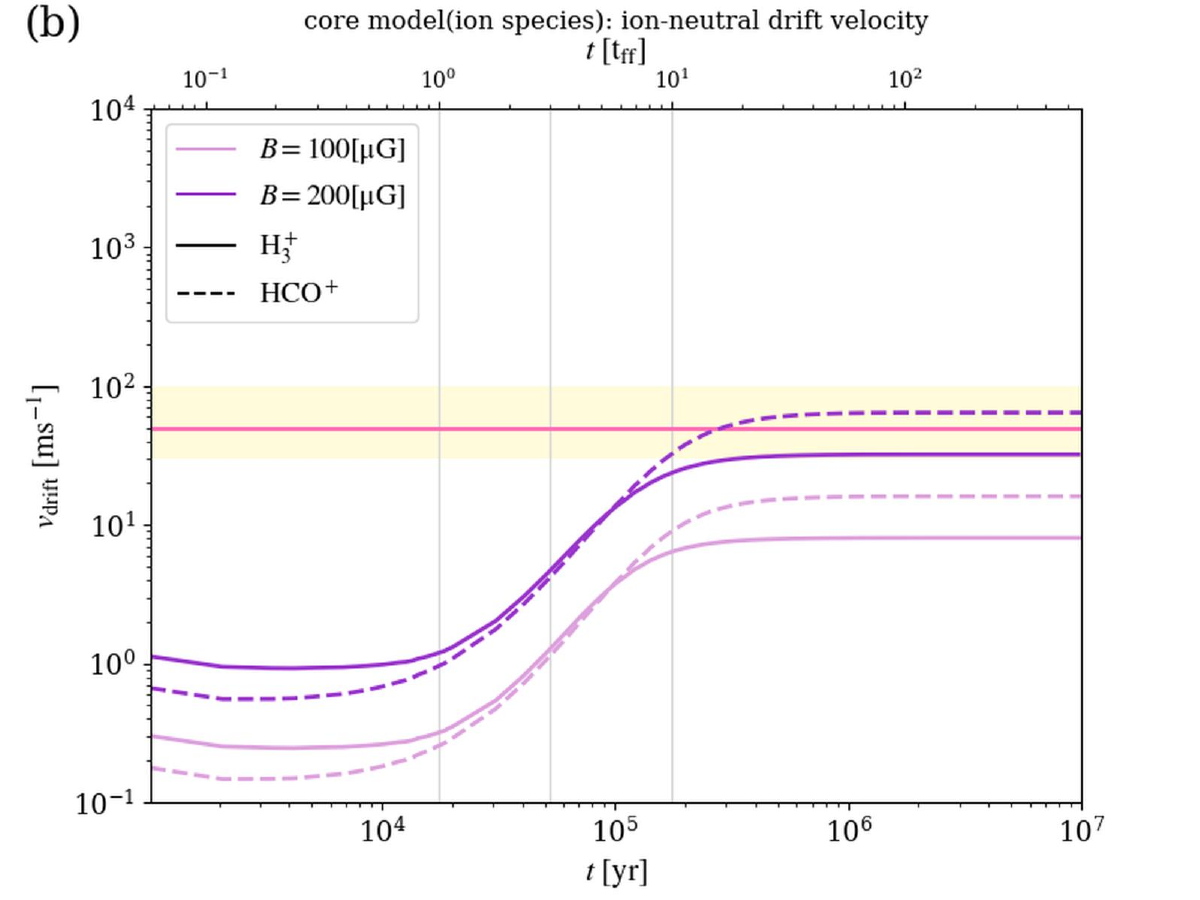}
\\[1ex]
\includegraphics[width=0.495\linewidth]{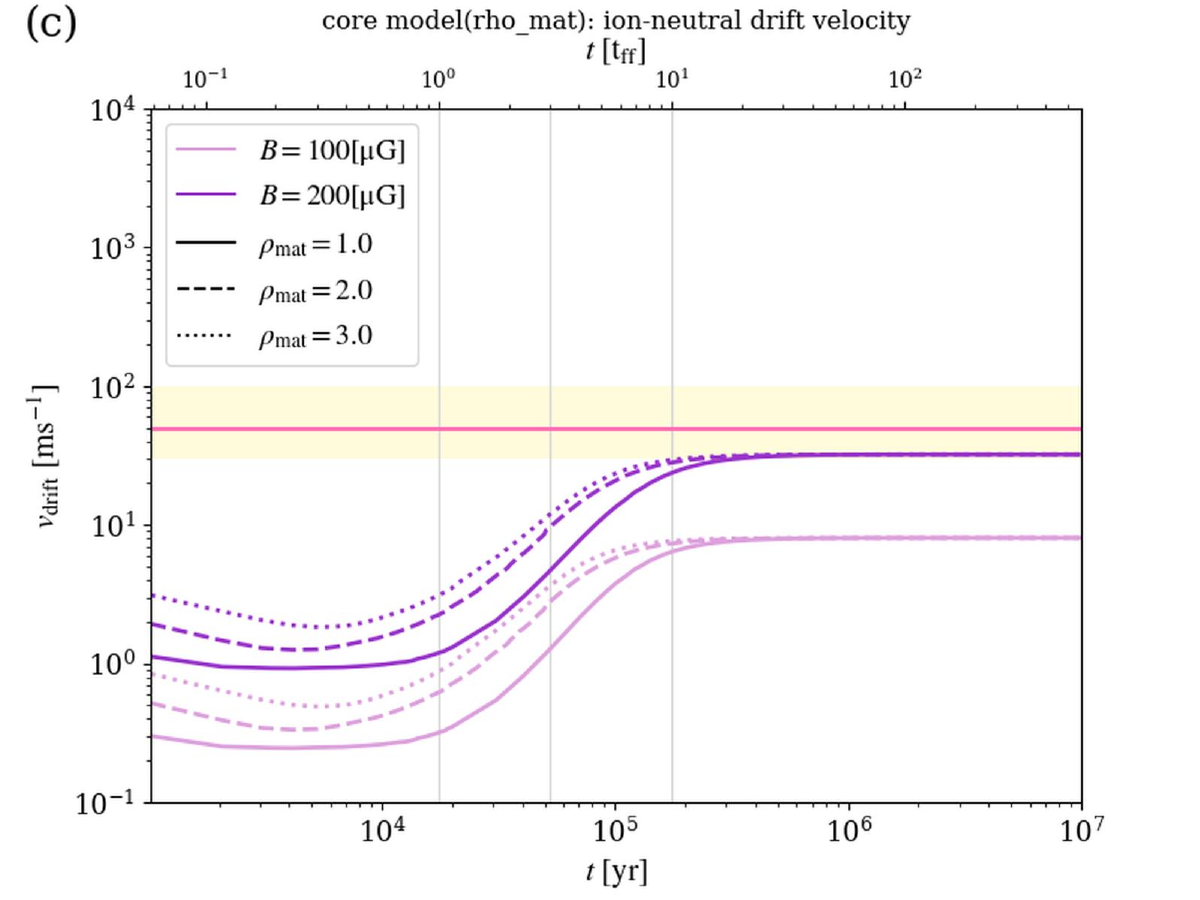}
\includegraphics[width=0.495\linewidth]{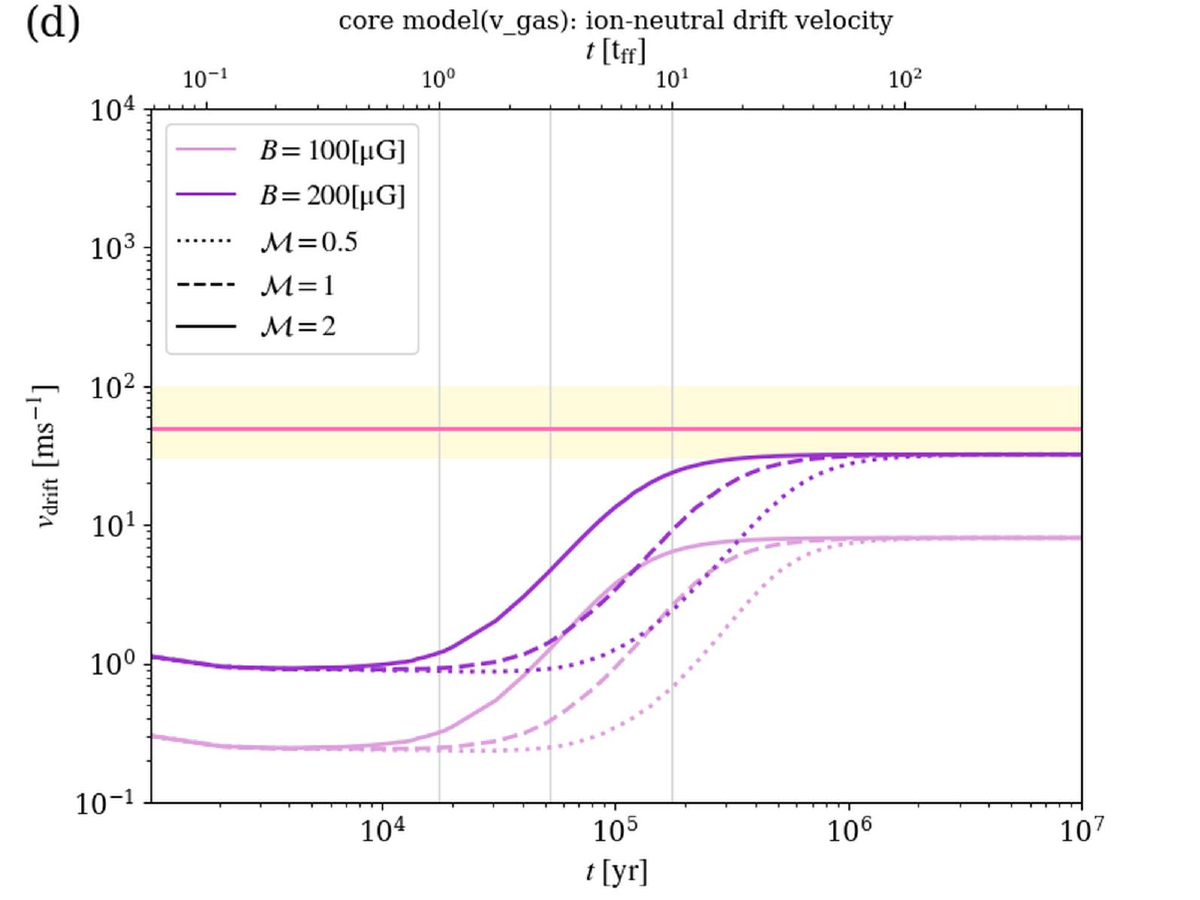}
\caption{Dependency of estimated $v_{\rm drift}$ on (a)the cosmic ray ionization rate; $\zeta_{\rm CR}$ (b)the dominant ion species; $s_{\rm ion}$ in gas phase (c)the intrinsic grain material density; $\rho_{\rm mat}$ (d)the gas velocity; $v_{\rm gas}$, in cloud core model. The difference in each parameter is indicated by the line styles.}
\label{fig:core_vdrift_zeta_ion_MatDens_vg}
\end{figure*}

Figure~\ref{fig:core_vdrift_zeta_ion_MatDens_vg} 
summarizes the dependence of \(v_{\rm drift}\) on four parameters:
(a) the cosmic-ray ionization rate \(\zeta_{\rm CR}\),
(b) the dominant ion species $s_{\rm ion}$,
(c) the grain intrinsic (material) density \(\rho_{\rm mat}\), and
(d) the gas (turbulent) velocity \(v_{\rm gas}\).
Line styles distinguish parameter values as indicated by the legend in each panel.

\paragraph{Cosmic-ray ionization rate.}
In panel~(a), variations in \(\zeta_{\rm CR}\) have little effect at early times (\(t \lesssim 10^{5}\,\mathrm{yr}\)), when the adsorption of charged particles onto the grains regulates the ionization degree.
By contrast, at later times (\(t \gtrsim 10^{5}\,\mathrm{yr}\)), dust grains grow sufficiently and gas-phase recombination dominates the loss of charged particles. 
In this regime, a larger \(\zeta_{\rm CR}\) maintains a higher ionization degree, yielding a lower resistivity and therefore a smaller \(v_{\rm drift}\).

\paragraph{Ion species.}
Panel~(b) shows how the choice of dominant ion species affects \(v_{\rm drift}\).
The dominant ion species in gas phase affects both the adsorption efficiency onto grains and gas-phase recombination rate.
Lighter ions, such as \(\mathrm{H_3^+}\), have larger thermal velocities, leading to more frequent collisions with dust grains and more efficient adsorption. 
This lowers the gas-phase ionization degree and consequently tends to increase \(v_{\rm drift}\) at early times.
On the other hand, in the later phase where dust growth has proceeded and gas-phase recombination dominates the ionization balance,
the lower recombination rate of \(\mathrm{H_3^+}\) maintains a higher ionization degree more efficiently than \(\mathrm{HCO^+}\).
As a result, \(v_{\rm drift}\) becomes smaller with \(\mathrm{H_3^+}\) at late times.

\paragraph{Grain material density.}
Under a fixed dust-to-gas mass ratio, variations in the intrinsic grain material density \(\rho_{\rm mat}\) change the normalization of the initial size distribution and hence the total grain surface area.
With increasing $\rho_{\rm mat}$, the overall number of grains decreases, resulting in a smaller total dust surface area.
Consequently, the number of very small charged grains that contribute to the conductivity also decreases.
This enhances the ambipolar resistivity and leads to a larger drift velocity (Figure~\ref{fig:core_vdrift_zeta_ion_MatDens_vg}(c)).

\paragraph{Gas velocity.}
Increasing \(v_{\rm gas}\) (i.e., the Mach number) raises grain–grain relative velocities and accelerates their coagulation, thereby shortening the timescale over which the \(v_{\rm drift}\) evolve.
We discuss these timescales in more detail in Section~\ref{sec:discusion}.\\

Taken together, the core-scale results indicate that reproducing the observed \(v_{\rm drift}\) at core densities requires both substantial dust growth and a magnetic field strength of order \(10^{2}\,\mu\mathrm{G}\) (specifically \(B\simeq 200\,\mu\mathrm{G}\) in our fiducial core model).
This conclusion holds across the explored ranges of \(\zeta_{\rm CR}\), dominant ion species $s_{\rm ion}$, grain material density \(\rho_{\rm mat}\), and turbulent velocity \(v_{\rm gas}\).

This naturally prompts the question of whether, at the lower densities (i.e., molecular clouds), measurements of \(v_{\rm drift}\) can likewise yield meaningful constraints on dust growth and on magnetic field strength.
In the cloud density, however, we conclude that dust growth is not always required.

\begin{figure*}[ht!]
\centering
\includegraphics[width=0.495\linewidth]{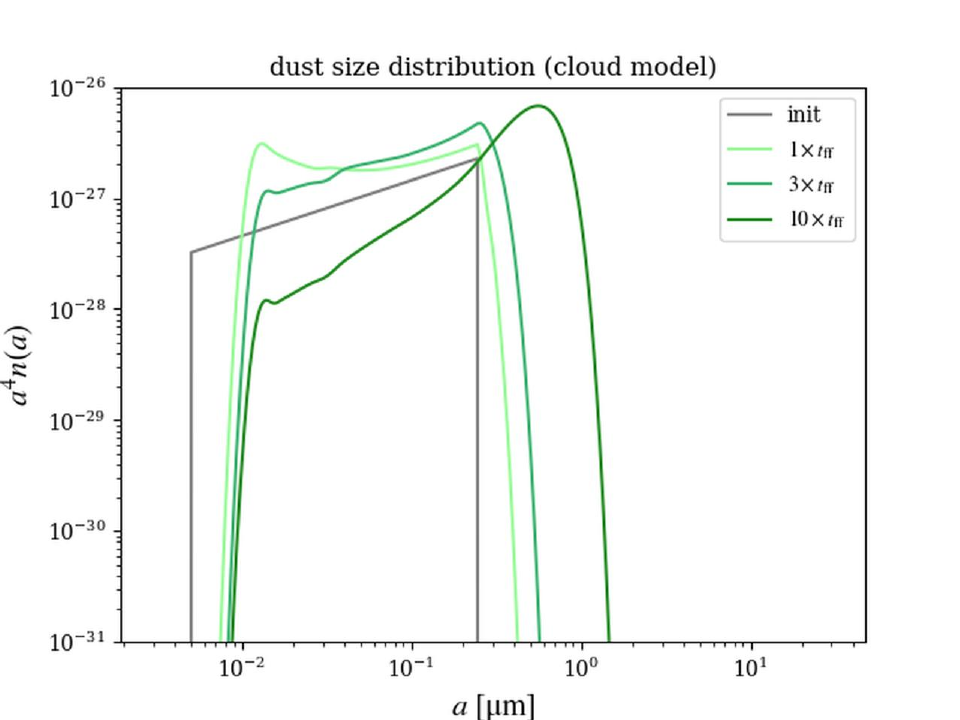}
\includegraphics[width=0.495\linewidth]{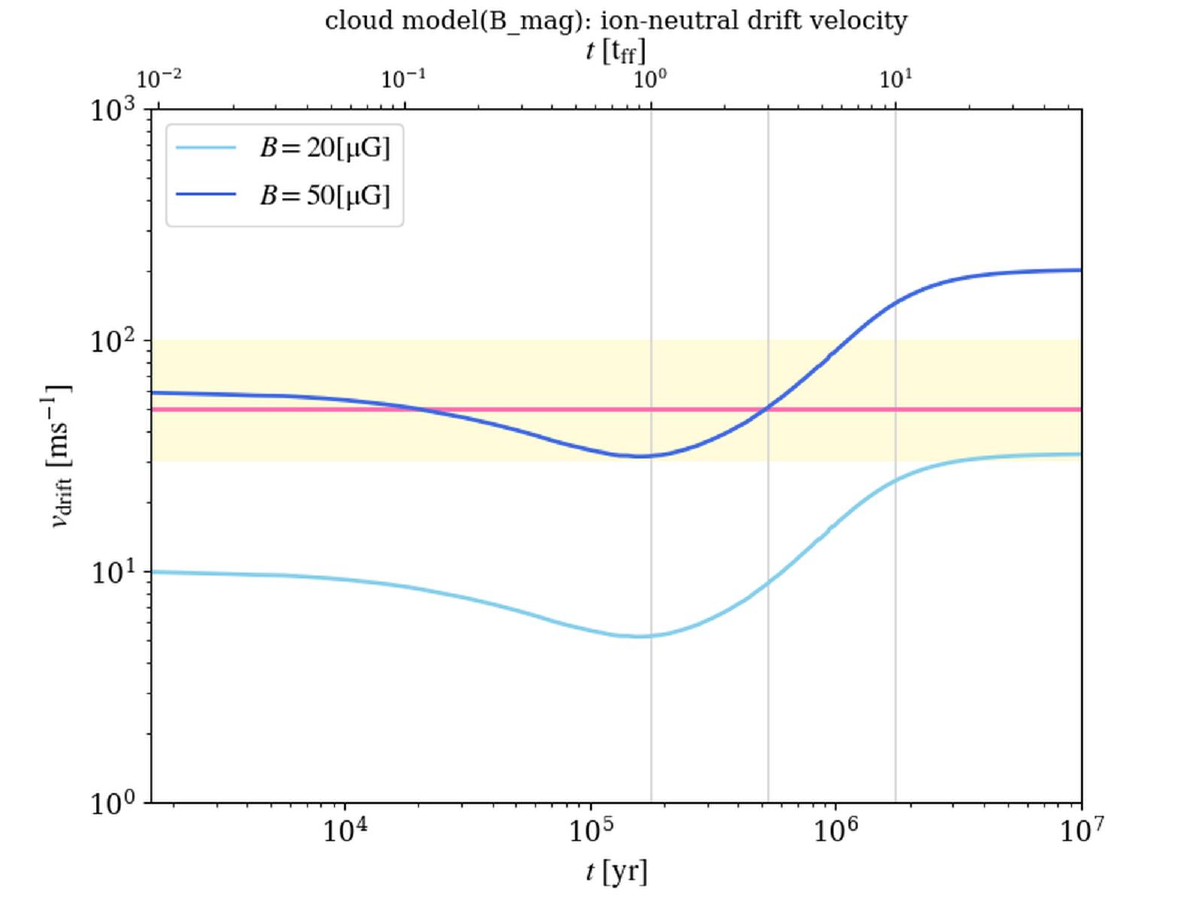}
\caption{Time evolution of dust size distribution(left) and ion-neutral drift velocity(right) for cloud model, shown in the same format as Figure \ref{fig:core_vdrift_Bmag}.
The light-blue and blue lines represent magnetic field strengths of $20\ \rm \mu G$ and $50\ \rm \mu G$, respectively.}
\label{fig:cloud_vdrift_Bmag}
\end{figure*}

The left panel of Figure~\ref{fig:cloud_vdrift_Bmag} shows the evolution of the grain-size distribution in the cloud model (its number density is $n_{\rm gas}=10^{4}\ {\rm cm^{-3}}$).
As in the core model, accretion rapidly depletes the small grains (\(a\lesssim 10\,\mathrm{nm}\)) at early phase, after which coagulation produces grains up to \(\sim 1\,\mu\mathrm{m}\).
A key difference from the core model is that the maximum grain size remains \(\sim 1\,\mu\mathrm{m}\) even by \(t \simeq 10\,t_{\rm ff}\).

\begin{figure*}[ht!]
\centering
\includegraphics[width=0.495\linewidth]{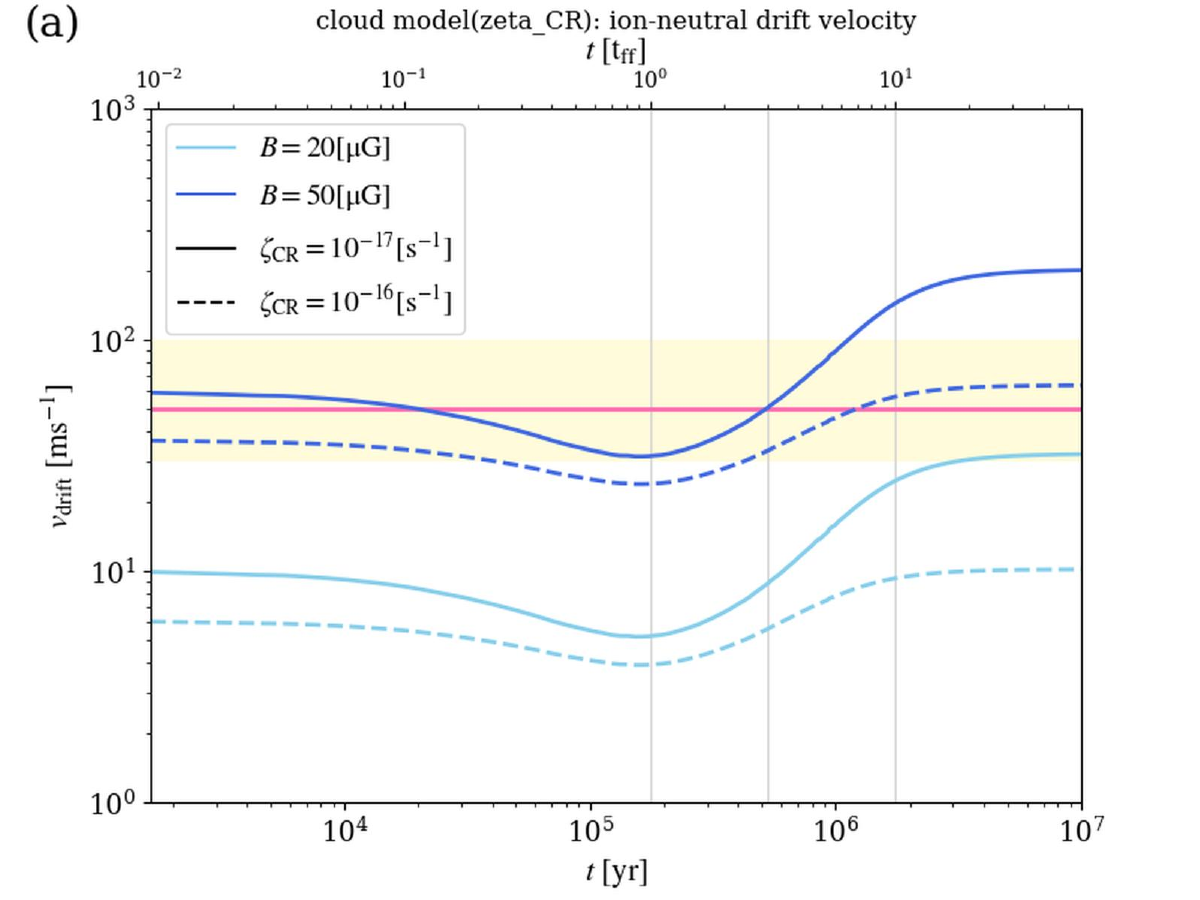}
\includegraphics[width=0.495\linewidth]{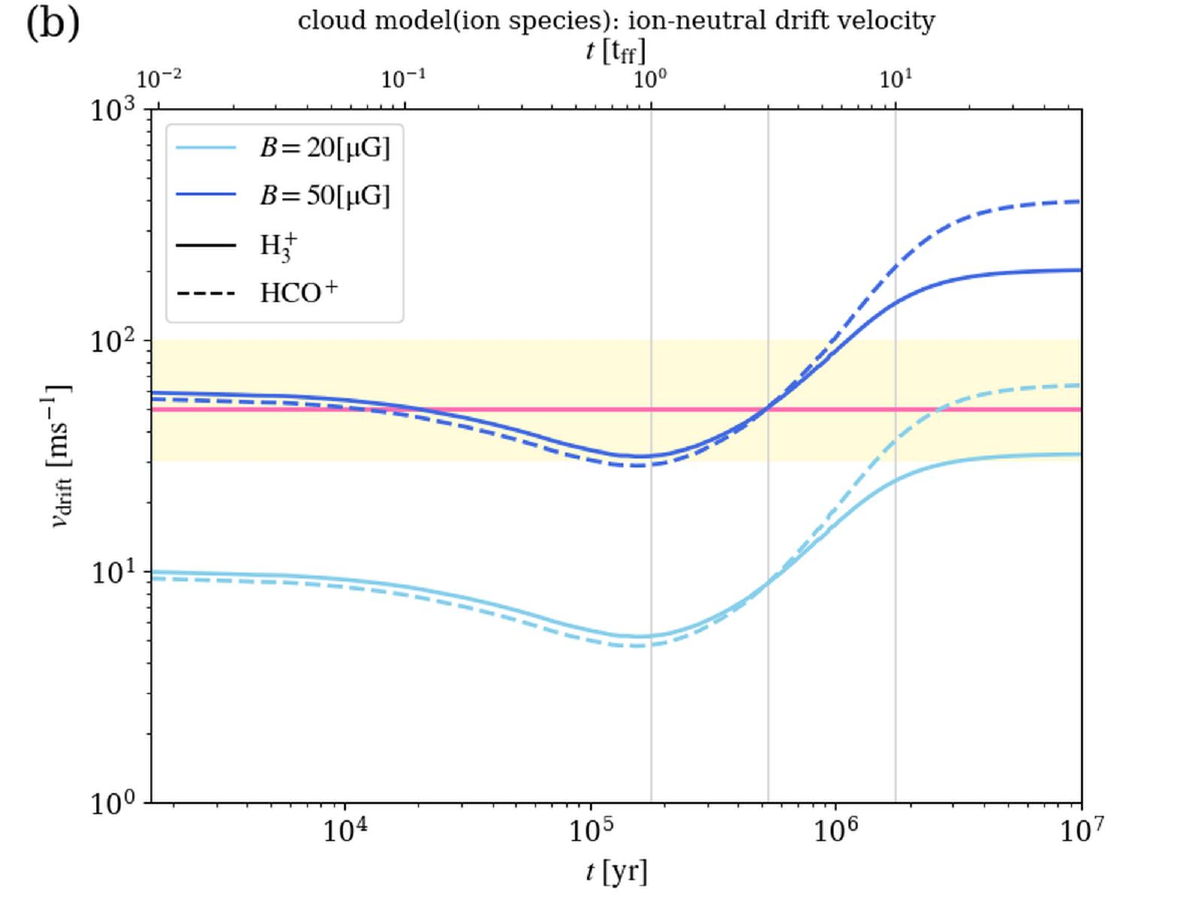}
\\[1ex]
\includegraphics[width=0.495\linewidth]{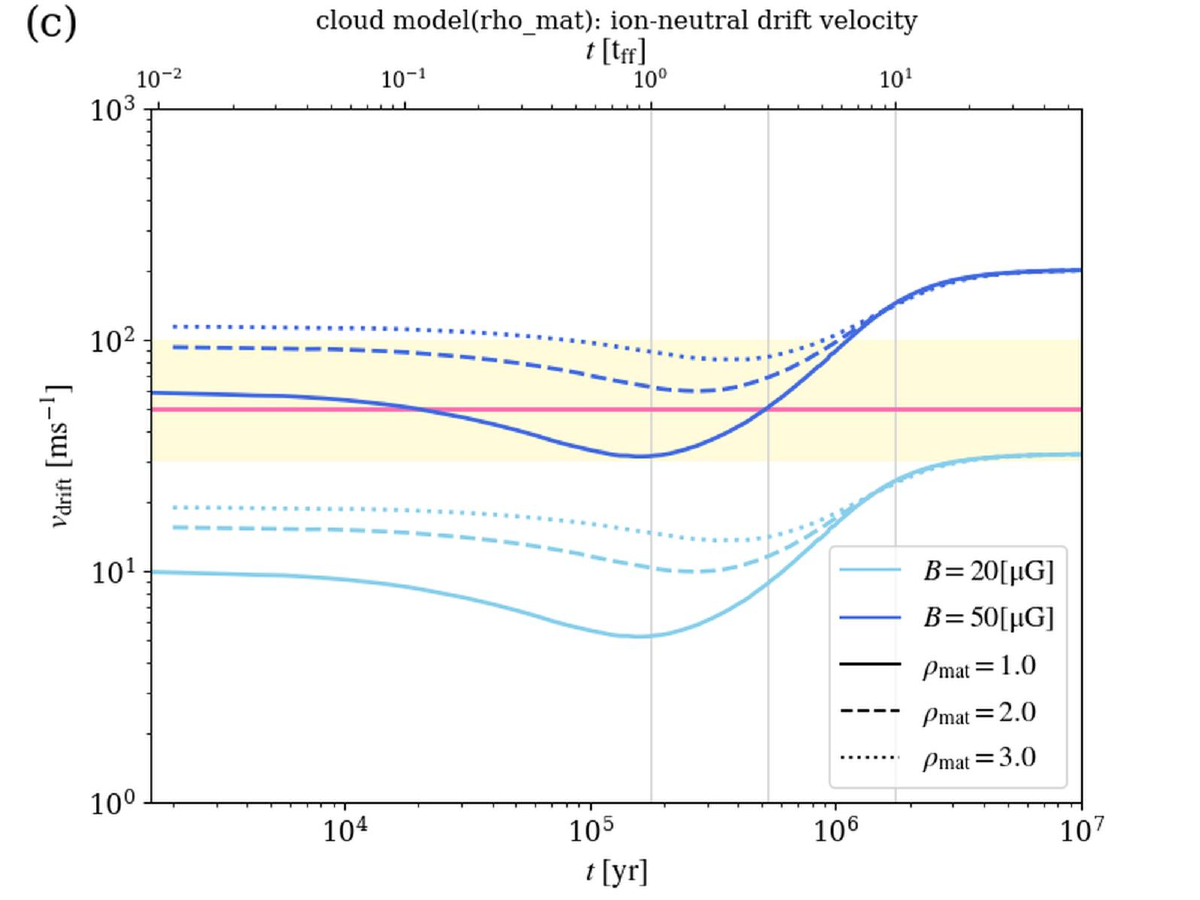}
\includegraphics[width=0.495\linewidth]{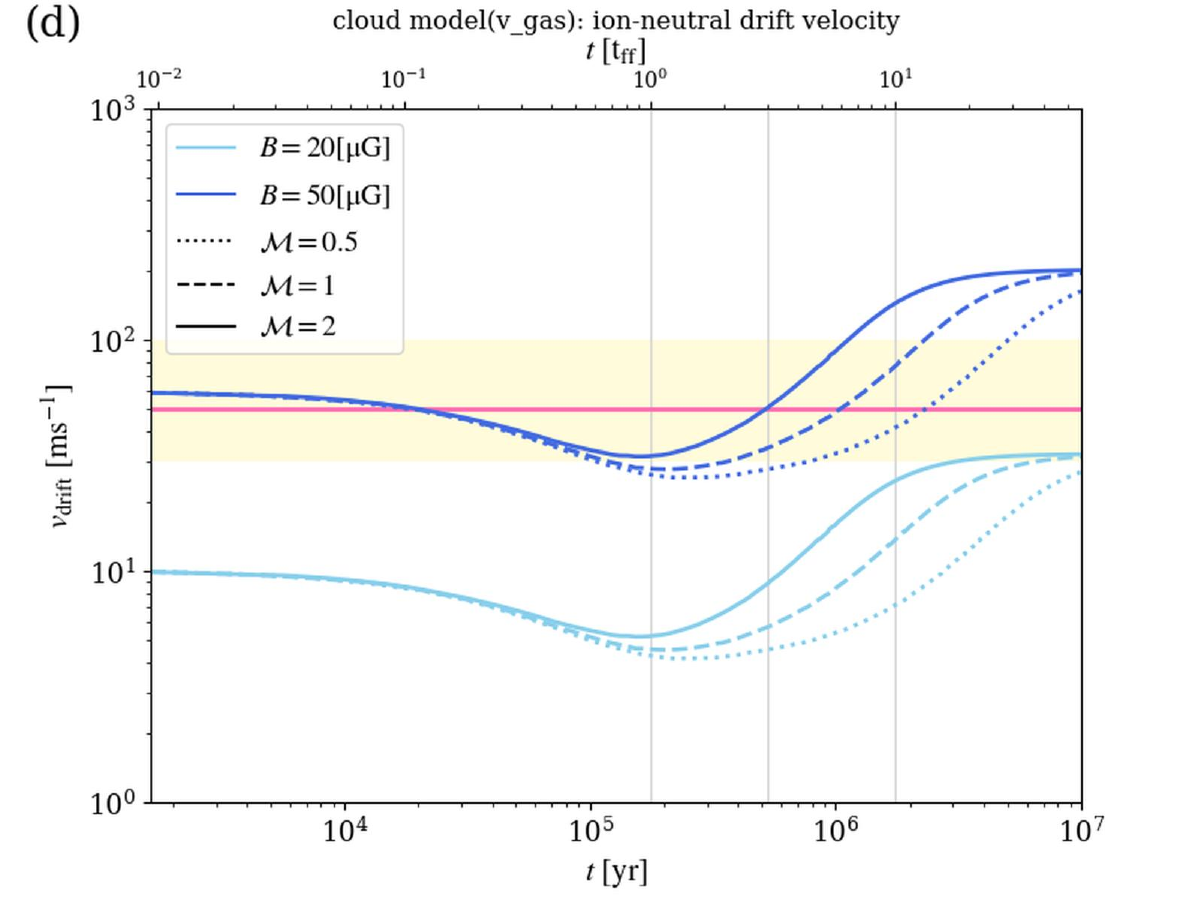}
\caption{The dependence of the time evolution of $v_{\rm drift}$ in the cloud model on (a)~$\zeta_{\rm CR}$ (b)~$s_{\rm ion}$ (c)~$\rho_{\rm mat}$ (d)~$v_{\rm gas}$, as in Figure~\ref{fig:core_vdrift_zeta_ion_MatDens_vg}.}
\label{fig:cloud_vdrift_zeta_ion_MatDens_vg}
\end{figure*}

Right panel shows the time evolution of the drift velocity. 
No single set of conditions uniquely reproduces the observed \(v_{\rm drift}\) in this model: for \(B=20~\mu\mathrm{G}\) the observed range is reached only after dust growth proceeds, whereas for \(B=50~\mu\mathrm{G}\) it is already achieved in the early phase without significant dust growth. 
Both values of magnetic field strength are acceptable for the values of molecular clouds \citep{2012ARA&A..50...29C}.
Figure~\ref{fig:cloud_vdrift_zeta_ion_MatDens_vg} illustrates the evolution of \(v_{\rm drift}\) in the cloud model and its dependence on the four parameters, shown in panels ((a)\(\zeta_{\rm CR}\), (b)$s_{\rm ion}$, (c)\(\rho_{\rm mat}\), and (d)\(v_{\rm gas}\)).
For each parameter, the qualitative trends are similar to those found in the core model (Figure~\ref{fig:core_vdrift_zeta_ion_MatDens_vg}).
Consequently, at cloud densities,  \(v_{\rm drift}\) alone does not tightly constrain the magnetic field strength because of degeneracies with dust evolution.

\section{DISCUSSION} \label{sec:discusion}
We have estimated ion-neutral drift velocity in molecular clouds and dense cores by coupling dust size evolution driven by accretion and coagulation with an ionization chemistry that self-consistently accounts for dust charging \citep[e.g.,][]{1987ApJ...320..803D,2022ApJ...934...88T}.
We then examine how the observed ion–neutral drift velocity, $v_{\rm drift}\sim 100~\mathrm{m\,s^{-1}}$, constrains the dust size distribution, the cosmic-ray ionization rate, and the magnetic field strength.
Our results indicate that dust growth plays a key role in reproducing the observed drift velocity, especially in cloud cores.

At core densities $(n_{\rm gas}\sim 10^{6}\,\mathrm{cm^{-3}}$), our models obtain $v_{\rm drift}\sim 100~\mathrm{m\,s^{-1}}$ only when 
(i) the grains have undergone substantial growth, leading to a strong reduction in total surface area (with the mass-weighted sizes extending to \(\sim 10~\mu\mathrm{m}\));
(ii) the magnetic field strength is on the order of $B\sim 200~\mu\mathrm{G}$; 
and (iii) cosmic-ray ionization rate is $\zeta_{\rm CR}\sim 10^{-17}\,\mathrm{s^{-1}}$.

In the absence of grain growth (i.e., an MRN-like abundance of small grains),  $v_{\rm drift}\sim 100~\mathrm{m\,s^{-1}}$ would instead require $B\gtrsim 1\,\mathrm{mG}$, significantly larger than typical estimates for dense cores \citep{2012ARA&A..50...29C}.
These trends follow directly from the sensitivity of the ambipolar diffusivity to the abundance of small grains, which contribute to the electrical conductivity as charge carriers \citep{2021MNRAS.505.5142Z,2022ApJ...934...88T}.

At cloud densities (\(n_{\rm gas}\sim 10^{4}\,\mathrm{cm^{-3}}\)), the observed range of \(v_{\rm drift}\) can be attained either with modest magnetic field strength once grain growth proceeds or with slightly stronger fields in the absence of grain growth.
Therefore, \(v_{\rm drift}\) alone does not provide strong constraint on magnetic field strength at cloud scales, because of its degeneracies with the dust size and magnetic field strength (Figure \ref{fig:cloud_vdrift_Bmag}, \ref{fig:cloud_vdrift_zeta_ion_MatDens_vg}).

Given that our core models place relatively tight constraints on the combination of magnetic field strength, cosmic-ray ionization rate, and dust growth required to reach \(v_{\rm drift}\sim 100~\mathrm{m\,s^{-1}}\), it is instructive to compare these conditions with those inferred for the well-studied starless (prestellar) core L1544 in Taurus.
\citet{2002ApJ...569..815T} estimated a central $\rm H_2$ number density of \(n_{\rm H_2}\simeq 1.4\times10^{6}\,\mathrm{cm^{-3}}\), with a characteristic flat inner radius of \(r\simeq 2.9\times10^{3}\,\mathrm{AU}\). 
Submillimeter polarization observations analyzed with the Chandrasekhar-Fermi method yield a plane-of-sky magnetic field strength of order \(B_{\rm pos}\sim 100\text{--}400~\mu\mathrm{G}\) on similar scales \citep{2004ApJ...600..279C}.
The cosmic-ray ionization rate in L1544 has been estimated to be
$\zeta_{\rm CR} \simeq (2\text{--}3)\times10^{-17}\,{\rm s^{-1}}$
\citep{2020MNRAS.495L...7B, 2021A&A...656A.109R}. 
These values are broadly comparable to the parameter space explored in our core models (see Table \ref{tab_param}).
Therefore, if the ion-neutral drift is observed in L1544 for example, particularly at the level discussed above, it lends support to the relevance of dust–growth–induced ambipolar diffusion for real prestellar cores.

In addition to the agreement in these profiles, it is worth noting that the L1544 core is known to exhibit the ''coreshine'' phenomenon \citep[][]{2010Sci...329.1622P}
which is interpreted as evidence for the presence of micron-sized dust grains.
This observational indication of dust growth provides independent support for the grain growth predicted in our model.

\begin{figure}[h]
    \centering
    \includegraphics[width=1.0\linewidth]{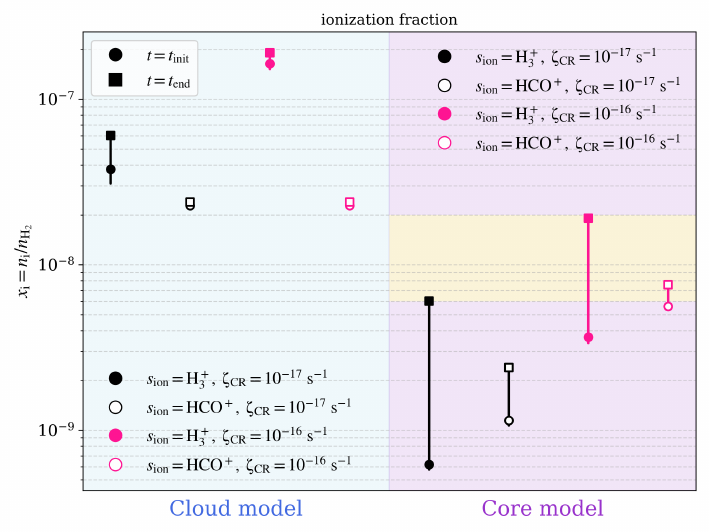}
    \caption{Ionization fraction $n_{\rm i}/n_{\rm H_2}$ in the molecular cloud model ($n_{\rm gas}=10^{4}\,{\rm cm^{-3}}$; blue-shaded region on the left) and the molecular cloud core model ($n_{\rm gas}=10^{6}\,{\rm cm^{-3}}$; purple-shaded region on the right).
    Circles and squares indicate the values at $t=t_{\rm init}$ and $t=t_{\rm end}$, respectively.
    Filled symbols correspond to $s_{\rm ion}={\rm H_3^+}$, while open symbols correspond to $s_{\rm ion}={\rm HCO^+}$.
    Black and pink symbols represent $\zeta_{\rm CR}=10^{-17}\,\rm{s^{-1}}$ and $\zeta_{\rm CR}=10^{-16}\,\rm{s^{-1}}$, respectively.
    The yellow shaded region indicates a factor-of-two range around the observed value $1.1\times10^{-8}$ reported for the L1544 core \citep{2002ApJ...565..344C}.}
    \label{fig:ionization_fraction}
\end{figure}

We next consider the ionization fraction.
For the L1544 core, the ionization fraction has been estimated to be \(x_i \equiv n_i/n_{\rm H_2} \simeq 1.1\times10^{-8}\) \citep{2002ApJ...565..344C}.
Are our results consistent with this value, and under what conditions can such consistency be achieved?
Figure~\ref{fig:ionization_fraction} shows the ionization fraction from our simulations.
The ionization fraction \(x_{\rm i}\) is calculated as
\begin{equation}
    x_{\rm i} = \frac{n_{\rm i}}{n_{\rm H_2}},
\end{equation}
where \(n_{\rm H_2} = (\mu/\mu_{\rm H_2})\,n_{\rm gas}\).
Here, \(\mu = 2.3\) and \(\mu_{\rm H_2} = 2.8\) are the mean molecular weights of the neutral gas and hydrogen gas, respectively.
In the cloud model, \(x_{\rm i}\) is in the range \(10^{-8}\text{--}10^{-7}\), whereas in the core model it is in the range \(\sim10^{-10}\text{--}10^{-8}\).
In both models, the ionization fraction increases as the dust grows.
These values are consistent with classical theoretical predictions \citep[e.g.,][]{1991ApJ...368..181N}.
More specifically, the values obtained with the ion species \(s_{\rm ion} = \mathrm{H_3^+}\) and cosmic-ray ionization rate $\zeta_{\rm CR}=1.0\times 10^{-16}\,\rm{s^{-1}}$ after sufficient dust growth are in good agreement with the ionization fraction in the L1544 core reported by \citet{2002ApJ...565..344C}.
This suggests that dust growth may play a key role in producing the low ionization fraction observed in the L1544 core.

Our cloud and core models employ a one-zone treatment that we regard as a useful benchmark rather than a precise description of real clouds or cores; it neglects spatial structure, line-of-sight averaging, and the full dynamical evolution from cloud to core.
We acknowledge that a number of studies have emphasized that extracting a reliable ion–neutral drift from line profiles is intrinsically challenging.
In order for \(v_{\rm drift}\) to be robustly inferred, the ion and neutral tracers must sample the same gas volume and be chemically co-evolving; otherwise, differences in critical densities or abundances can mimic or mask true drift \citep[e.g.,][]{2008ApJ...677.1151L, 2012ApJ...760...57T}.
Line radiative transfer and projection further complicate the interpretation: line-of-sight integration, magnetic-field geometry, and turbulent fluctuations can reduce the observable signature of a given 3D drift, so that even a true subsonic drift may appear at a reduced level in the observed linewidths \citep[e.g.,][]{2010MNRAS.406.1201T,2014MNRAS.438..663H}.
More recently, \citet{2023MNRAS.521.5087T} combined chemodynamical non-ideal MHD simulations and non-LTE radiative transfer for the simulation results, and showed that secure detection of ion–neutral drift generally requires stringent observing conditions and careful tracer selection.
These works underscore that currently reported values of \(v_{\rm drift}\) carry systematic uncertainties of a factor of a few that must be borne in mind when comparing with models.

Within our idealized one-zone framework, however, we find that dust growth alone can change the predicted \(v_{\rm drift}\) by nearly an order of magnitude at fixed density $n_{\rm gas}$, magnetic field strength \(B\), and cosmic-ray ionization rate \(\zeta_{\rm CR}\).
Thus, even if the absolute observationally inferred drift velocities are revised downward by geometry or radiative-transfer effects, the qualitative conclusion that the dust size distribution has a strong impact on \(v_{\rm drift}\) remains robust.

Our one-zone treatment with fixed  $n_{\rm gas}$ and $B$ also has a limitation in evaluating the dust growth timescale.
In reality, dust coagulation proceeds within a time-dependent cloud-to-core evolution, which involves the dynamics of gas density (free-fall time) and magnetic field. 
As we have seen, the coagulation timescale to reach micron sizes at \(n_{\rm gas}\sim 10^{4}\text{-}10^{6}\,\mathrm{cm^{-3}}\) exceeds a free-fall time, whereas recent simulations suggest that filament-to-core formation proceeds over several free-fall times ($\sim10\ \rm\mu G$ at $n_{\rm gas}=10^4 \ \rm cm^{-3}$) \citep{2024ApJ...963..106M, 2025PASJ...77..277F}.

These challenges motivate fully coupled 3D non-ideal MHD calculations that evolve gas density, magnetic fields, chemistry, and dust grain simultaneously to identify when and where efficient grain growth occurs during cloud contraction.
Furthermore, combining time-dependent multi-dimensional non-ideal MHD and forward modelling of line emission will enable the dust microphysics and observational systems to be treated as complementary components on equal footing in forming the measured ion-neutral drift.

In addition, our treatment of dust size evolution can be extended in several ways.
Our coagulation model employs a relative-velocity prescription that includes only Brownian and turbulence-driven components.
However, several additional processes can modify the dust size evolution.
For example, we neglect detailed collision physics such as electrostatic repulsion: Coulomb barriers between like-charged grains can inhibit sticking at small sizes \citep[e.g.,][]{2009ApJ...698.1122O}.
In magnetized clouds, the small charged grains can become magnetically tied to field lines and participate in ambipolar drift together with ions \citep[e.g.,][]{1996ApJ...468..749C, 2020A&A...643A..17G, 2020A&A...641A..39S}.
Quantifying the combined impact of these effects within a framework that self-consistently incorporates dust charging and ambipolar diffusion will require more comprehensive dust-kinetics models and constitutes an important direction for future work.

Although these future challenges remain, we think that the observation of ion-neutral drift velocity is a promising diagnostic for constraining the dust size distribution and magnetic field strength in molecular cloud cores.

\begin{acknowledgments}
This study was supported by the JST FOREST Program, Grant Number JPMJFR2234.
We sincerely thank all those whose valuable discussions, whether in our daily work or at conferences and workshops, have contributed to the development of this study.
In particular, we are grateful to Dr. Shinsuke Takasao, who kindly joined our discussions and made them more rewarding.
We further wish to express our thanks to Dr. Silvia Spezzano, who participated in our discussions and warmly contributed to a pleasant and fruitful exchange.
Finally we thank the anonymous referee for constructive comments and suggestions that greatly improved the manuscript.
\end{acknowledgments}

\clearpage
\bibliography{bib}{}
\bibliographystyle{aasjournal}



\end{document}